\documentclass[showpacs,preprintnumbers,amsmath,amssymb]{revtex4}
\usepackage{graphicx}
\usepackage{dcolumn}
\usepackage{bm}
\usepackage{epsfig}

\newcommand{\dhd}{{\textstyle d}
\lower.03ex\hbox{\kern-0.40em$^{\scriptstyle-}$}\kern-0.08em{}}

\newcommand{\halo}{\hat{\cal O}}

\newcommand{\half}{{1\over 2}}
\newcommand{\bu}{{\bullet}}

\newcommand{\barz}{{\bar z}}

\newcommand{\calr}{{\cal R}}  
\newcommand{\calo}{{\cal O}}  
 
\newcommand{\calu}{{\cal U}} 
\newcommand{\calv}{{\cal V}} 
\newcommand{\calz}{{\cal Z}}

\begin{document}

\preprint{JLAB-THY-09-1110, CPHT-RR124.1109, LPT-ORSAY 09-105}

\title{High-energy amplitudes in ${\cal N}$=4 SYM in the next-to-leading order}

\author{Ian Balitsky }
\address{Physics Dept., ODU, Norfolk VA 23529
and \\
Theory Group, Jlab, 12000 Jefferson Ave, Newport News, VA 23606,USA\\
E-mail: balitsky@jlab.org}

\author{Giovanni A. Chirilli}
\address{Centre de Physique ThŽorique,
Ecole polytechnique, CNRS,
91128 Palaiseau, France and\\
LPT, Universit\'e Paris-Sud, CNRS, 91405 Orsay, France\\
E-mail: chirilli@cpht.polytechnique.fr}

\begin{abstract}
The high-energy behavior of the ${\cal N}$=4 SYM amplitudes in the Regge limit
 can be calculated order by order in perturbation theory 
using  the high-energy operator expansion in Wilson lines.
 At large $N_c$, a typical four-point amplitude  is determined by a single BFKL  pomeron. The conformal structure
of the four-point amplitude is fixed in terms of two functions: pomeron intercept and the coefficient function in front of the pomeron 
(the product of two residues).  The pomeron intercept is universal while the coefficient function
depends on the correlator in question.
The intercept is known in the first two orders in coupling constant: BFKL intercept and NLO BFKL intercept
calculated in Ref. \cite{lipkot}.  As an example of using the Wilson-line OPE, we calculate the coefficient
function in front of the pomeron for the correlator of four $Z^2$ currents in the first two orders in perturbation theory. 
\end{abstract}

\maketitle
\section{Introduction\label{aba:sec1}}

The high-energy scattering in a gauge theory can be described in terms of Wilson lines - infinite gauge
 factors ordered along the straight lines (see e.g. the review \cite{mobzor}). 
Indeed, the fast particle moves along its straight-line classical trajectory and the only quantum effect is 
the eikonal phase factor acquired along this propagation path. For a fast particle scattering 
off some target, this eikonal phase factor is a Wilson line - an infinite gauge link ordered along the straight 
line collinear to particle's velocity $n^\mu$:
\begin{equation}
U^Y(x_\perp)={\rm Pexp}\Big\{ig\int_{-\infty}^\infty\!\!  du ~n_\mu 
~A^\mu(un+x_\perp)\Big\},~~~~
\label{defU}
\end{equation}
Here $A_\mu$ is the gluon field of the target, $x_\perp$ is the transverse
position of the particle which remains unchanged throughout the collision, and the 
index $Y$ labels the rapidity of the particle.

The high-energy behavior of the amplitudes can be studied in the 
framework of the rapidity evolution of Wilson-line operators forming color dipoles \cite{mu94,nnn}.  The main idea is factorization in rapidity \cite{npb96,prld}: we separate 
a typical functional integral describing scattering of two particles into ({\bf i}) the integral over the gluon (and gluino) fields with rapidity close to the rapidity of the spectator $Y_A$,
({\bf ii}) the integral over the gluons with rapidity close to the rapidity of the target $Y_B$, and ({\bf iii}) the integral over the intermediate region of rapidities $Y_A>Y>Y_B$, see Fig. 1.
The result of the first integration is a certain coefficient function (``impact factor'') times color dipole (ordered in the direction of  spectator's velocity)
with rapidities up to $Y_A$. Similarly, the result of second integration is again the impact factor times color dipole ordered in the direction of target's velocity
with rapidities greater than $Y_B$. The result of last integration is the correlation function of two dipoles which can be calculated using the evolution 
equation for color dipoles which is known in the leading and next-to-leading order. 
As an example of practical use of this factorization scheme in the NLO approximation, in present paper we calculate the high-energy behavior 
of the ``scattering amplitude of scalar particles'' (the four-point correlation function of scalar currents).
\begin{figure}[htb]
\psfig{file=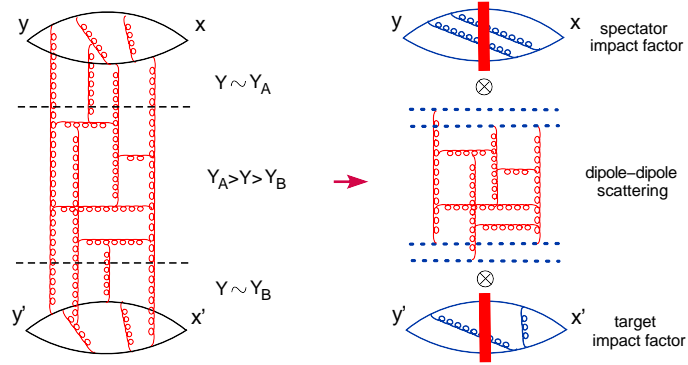,width=90mm}
\caption{High-energy factorization}
\label{aba:fig1}
\end{figure}

The high-energy (Regge) limit of a four-point amplitude $A(x,y;x',y')$ in the coordinate space can be achieved as 
\begin{eqnarray}
&&\hspace{-1mm}
x=\rho x_\ast p_1+x_\perp,~~~ y=\rho y_\ast p_2+y_\perp, ~~~~~~~
\nonumber\\
&&\hspace{-1mm}
x'=\rho' x_\bu p_2 +x'_\perp,~~~  y'=\rho' y'_\bu p_2 +y'_\perp
\label{reggelimit}
\end{eqnarray}
 with $\rho,\rho'\rightarrow\infty$ and $x_\ast>0>y_\ast$, $x'_\bu>0>y'_\bu$. (Strictly speaking, $\rho\rightarrow\infty$ or $\rho'\rightarrow \infty$ would be sufficient to reach the Regge limit). 
 Hereafter we use the notations $x_\bu=p_1^\mu x_\mu$, $x_\ast=p_2^\mu x_\mu$
where $p_1$ and $p_2$  are light-like vectors normalized by $2(p_1,p_2)=s$. These ``Sudakov variables'' are related to 
the usual light-cone coordinates
$x^\pm={1\over\sqrt{2}}(x^0\pm x^3)$ by $x_\ast=x^+\sqrt{s/2},~x_\bu=x^-\sqrt{s/2}$ so $x={2\over s}x_\ast p_1+{2\over s}x_\bu p_2+x_\perp$. We use the (1,-1,-1,-1) metric  so 
$x^2={4\over s}x_\bu x_\ast -\vec{x}_\perp^2$.
In the Regge limit (\ref{reggelimit}) the full conformal group  reduces to M\"{o}bius subgroup SL(2,C) leaving the transverse plane $(0,0,z_\perp)$ invariant. 

For simplicity, let us consider correlation function of four scalar currents
\begin{eqnarray}
(x-y)^4(x'-y')^4\langle \calo(x) \calo^\dagger(y)\calo(x')\calo^\dagger(y') \rangle~
\label{correl1}
\end{eqnarray}
where ${\calo}\equiv{4\pi^2\sqrt{2}\over \sqrt{N_c^2-1}}{\rm Tr\{Z^2\}}$ ($Z={1\over\sqrt{2}}(\phi_1+i\phi_2)$) is a renorm-invariant chiral primary operator.

In a conformal theory this four-point amplitude $A(x,y;x',y')$ depends on two conformal ratios which can be chosen as
\begin{eqnarray}
&&\hspace{-1mm}
R~=~{(x-x')^2(y-y')^2\over (x-y)^2(x'-y')^2},~~~~
\nonumber\\
&&\hspace{-1mm}r~=~R\Big[1-{(x-y')^2(y-x')^2\over (x-x')^2(y-y')^2}+{1\over R}\Big]^2
\label{cratios1}
\end{eqnarray}
In the Regge limit (\ref{reggelimit}) the conformal ratio $R$ scales as $\rho^2{\rho'}^2$ while $r$  does not depend on $\rho$ or $\rho'$. 

 As demonstrated in Ref. \cite{cornalba}, the pomeron contribution in a conformal theory can be represented as an integral over 
 one real variable $\nu$
\begin{eqnarray}
&&\hspace{-3mm}
(x-y)^4(x'-y')^4\langle \calo(x) \calo^\dagger(y)\calo(x')\calo^\dagger(y') \rangle~
\label{koppinkoop}\\
&&\hspace{-3mm}
=~{i\over 2}\!\int\! d\nu~\tilde{ f}_+(\nu)
{\tanh\pi\nu\over \nu}F(\nu)
\Omega(r,\nu)R^{\half\omega(\nu)}
\nonumber
\end{eqnarray}
Here $\omega(\nu)\equiv \omega(0,\nu)$ is the pomeron intercept, $\tilde{f}_+(\nu)\equiv \tilde{f}_+(\omega(\nu))$
where  $\tilde{f}_+(\omega)=(e^{i\pi\omega}-1)/\sin\pi\omega$ is the signature factor in the coordinate space, and
 $F(\nu)$ is the ``pomeron residue''  (strictly speaking, the product of two  pomeron residues). The conformal function $\Omega(r,\nu)$
is given by a hypergeometric function (see Ref. \cite{penecostalba}) but for our purposes it is convenient to use
 the representation in terms of the two-dimensional integral
\begin{eqnarray}
&&\hspace{-5mm}
\Omega(r,\nu)~=~{\nu^2\over\pi^3}
\!\int\! d^2z \Big[{\kappa^2\over (2\kappa\cdot\zeta)^2}\Big]^{\half +i\nu} \Big[{{\kappa'}^2\over (2\kappa'\cdot\zeta)^2}\Big]^{\half -i\nu}
\label{integral7}
\end{eqnarray}
where $\zeta\equiv p_1+{z_{\perp}^2\over s}p_2+z_{\perp}$ and
\begin{eqnarray}
&&\hspace{-5mm}
\kappa~=~{\sqrt{s}\over 2x_\ast}(p_1-{x^2\over s}p_2+x_\perp)-{\sqrt{s}\over 2y_\ast}(p_1-{y^2\over s}p_2+y_\perp)
\label{kappas}\\
&&\hspace{-5mm}
\kappa'~=~{\sqrt{s}\over 2x'_\bu}(p_1-{{x'}^2\over s}p_2+x'_\perp)-{\sqrt{s}\over 2y'_\bu}(p_1-{{y'}^2\over s}p_2+y'_\perp)
\nonumber
\end{eqnarray}
are  two SL(2,C)-invariant vectors \cite{penecostalba} 
such that 
\begin{equation}
\kappa^2{\kappa'}^2~=~{1\over R}~~~~~~{\rm and}~~~~~~~~4(\kappa\cdot\kappa')^2~=~{r\over R}
\label{Rr}
\end{equation}
Here $x^2=-x_\perp^2,~{x'}^2=-{x'}_\perp^2$ and similarly for $y$.  
Note that all the dependence on large energy 
($\equiv$ large $\rho,\rho'$) is contained in  $R^{\half\omega(\nu)}$.

The dynamical information about the conformal theory is encoded in two functions: pomeron intercept and pomeron residue.
The pomeron intercept is known both in the small and large $\alpha_s$ limit. At small $\alpha_s$ \cite{bfkl}
\begin{eqnarray}
&&\hspace{-2mm} 
\omega(\nu)~=~{\alpha_s\over \pi}N_c\Big[\chi(\nu)+{\alpha_sN_c\over 4\pi}\delta(\nu)
\Big],
\nonumber\\
&&\hspace{-1mm}
\delta(\nu)~=~
6\zeta(3)-
{\pi^2\over 3}\chi(\nu)+\chi''(\nu)
-~2\Phi(\nu)-2\Phi(-\nu)
\label{eigen1}
\end{eqnarray}
where $\chi(\nu)=2\psi(1)-\psi(\half +i\nu)-\psi(\half -i\nu)$ and \cite{lipkot}
\begin{eqnarray}
&&\hspace{-1mm}
\Phi(\nu)~=~-\int_0^1\!{dt\over 1+t}~t^{-\half+i\nu}
\Big[{\pi^2\over 6}+2{\rm Li}_2(t)\Big]
\label{fi}
\end{eqnarray}
Our main goal is the description of the amplitude in the next-to-leading order in perturbation theory, but 
it is worth noting that the pomeron intercept is known also in the limit of large 
't Hooft coupling $\lambda=4\pi\alpha_sN_c$
\begin{equation}
\hspace{-1mm}
\omega(\nu)~=~2-{\nu^2+4\over 2\sqrt{\lambda}}
\end{equation}
where 2 is the graviton spin and the first correction was calculated in Ref. \cite{klov,brtan}.

The pomeron residue $F(\nu)$ is known in the leading order both at small \cite{penecostalba, penecostalba2,penedones} and large \cite{cornalba} 't Hooft coupling
\begin{equation}
\hspace{-1mm}
F(\nu)~\stackrel{\lambda\rightarrow 0}{\rightarrow}~\lambda^2{\pi\sin\pi\nu \over 4\nu\cos^3\pi\nu},~~~~
F(\nu)~\stackrel{\lambda\rightarrow \infty}{\rightarrow}~{\pi^3\nu^2(1+\nu)^2\over \sinh^2\pi\nu}
\label{loif}
\end{equation}
To find the NLO amplitude, we must also calculate the ``pomeron residue'' $F(\nu)$
in the next-to-leading order. 
In the rest of the paper we will do this using the high-energy operator product expansion in Wilson lines \cite{npb96}.

\section{Operator expansion in conformal dipoles}
As we discussed above, the main idea behind the high-energy operator expansion is the rapidity factorization.  At the first step, we integrate 
over gluons with rapidities $Y>\eta$ and leave the integration over $Y<\eta$ for later time, see Fig. 2. 
\begin{figure}[htb]
\psfig{file=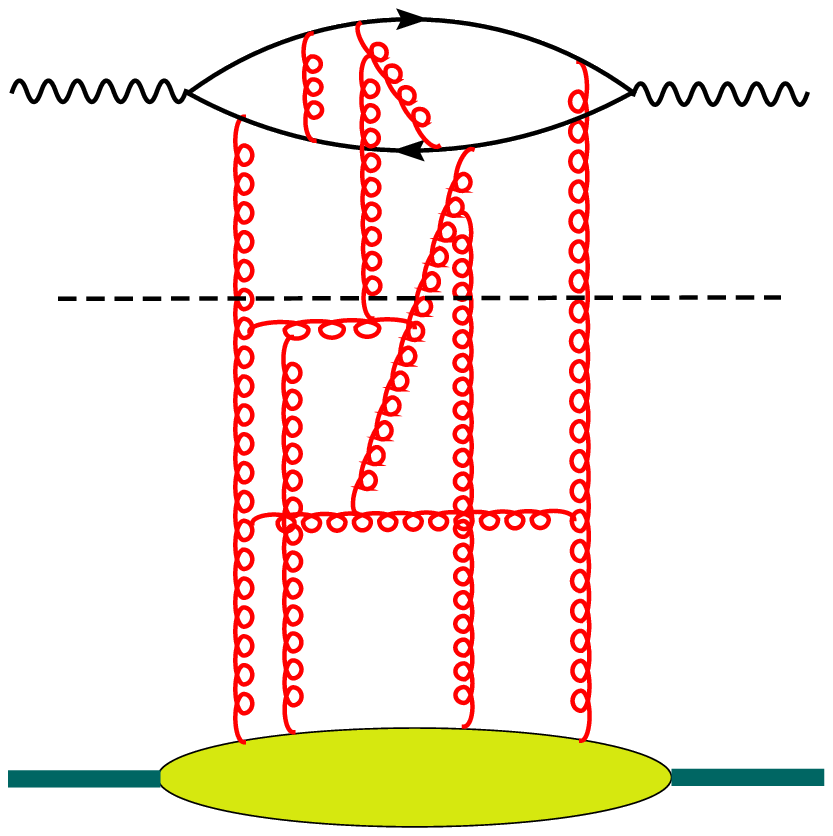,width=46mm} 
\hspace{2mm} 
\psfig{file=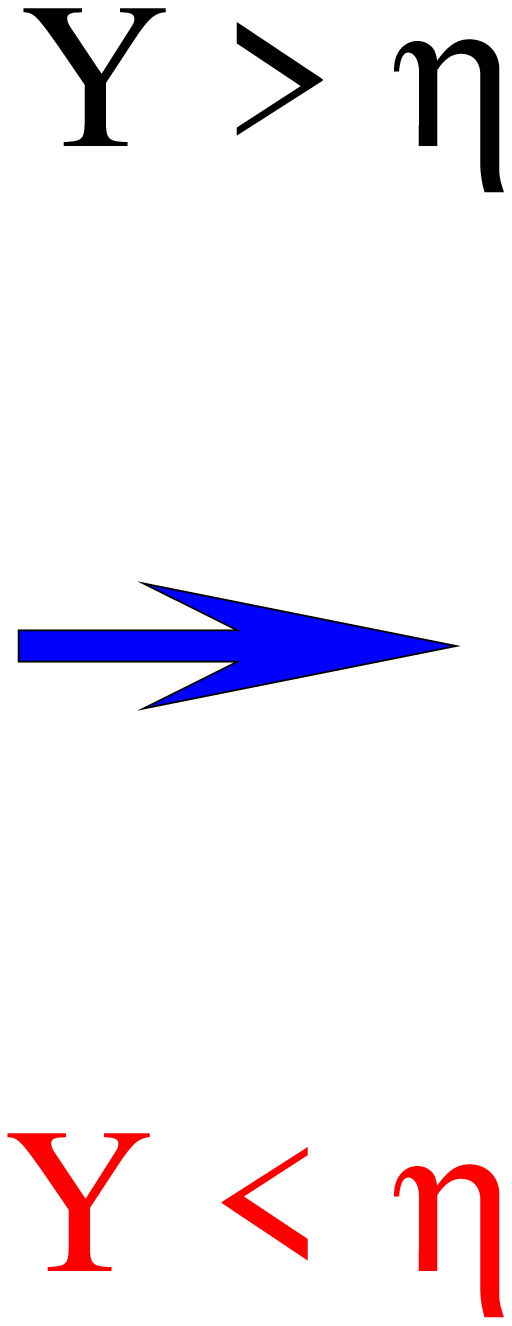,width=6mm} 
\hspace{2mm}
\psfig{file=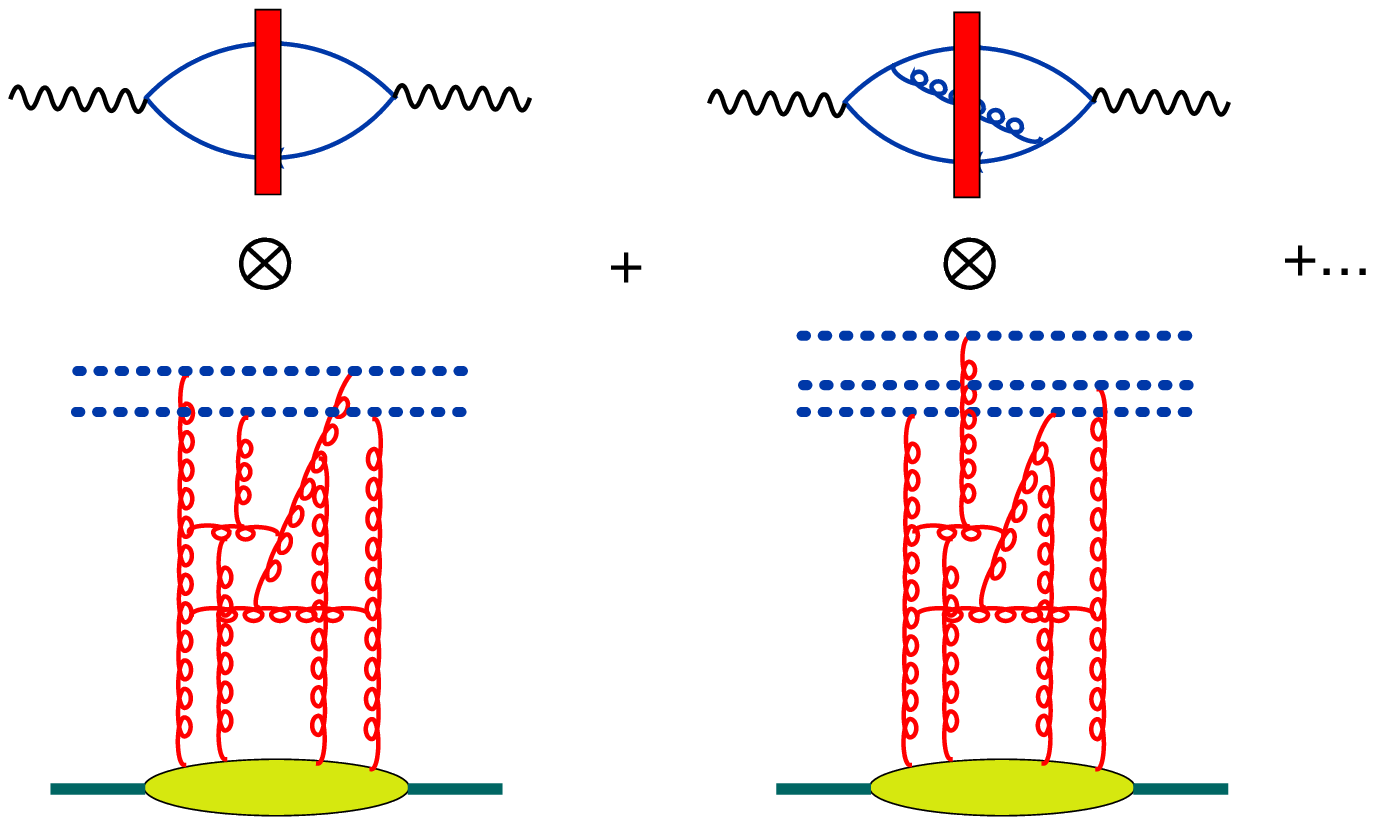,width=82mm}
\caption{High-energy operator expansion in Wilson lines}
\label{aba:fig2}
\end{figure}
The result of the integration is the coefficient
function (``impact factor'') in front of the Wilson-line operators with rapidities up to $\eta=\ln\sigma$:
\begin{eqnarray}
&&\hspace{-2mm} 
 U^\sigma_x~=~{\rm Pexp}\Big[ig\!\int_{-\infty}^\infty\!\! du ~p_1^\mu A^\sigma_\mu(up_1+x_\perp)\Big]
 \nonumber\\ 
&&\hspace{-2mm}
A^\sigma_\mu(x)~=~\int\!d^4 k ~\theta(\sigma-|\alpha_k|)e^{-ik\cdot x} A_\mu(k)
\label{cutoff}
\end{eqnarray}
For the $T$-product of scalar currents currents $\calo$ this coefficient function has the form \cite{nlobksym}:
\begin{eqnarray}
&&\hspace{-5mm}
(x-y)^4 T\{\halo(x)\halo^\dagger(y)\}~
=~{1\over \pi^2(N_c^2-1)}\!\int\! {d^2 z_1d^2 z_2\over z_{12}^4}~
\calr^2~\Big[{\rm Tr}\{\hat{U}^\sigma_{z_1}\hat{U}^{\dagger\sigma}_{z_2}\}
 \label{ope1}\\
&&\hspace{-5mm}
-~{\alpha_s\over\pi^2}\!\int\! d^2 z_3~{z_{12}^2\over z_{13}^2z_{23}^2}
\Big[\ln{ s\over 4}\sigma\calz_3-{i\pi\over 2}+C\Big]
[ {\rm Tr}\{T^n\hat{U}^\sigma_{z_1}\hat{U}^{\dagger\sigma}_{z_3}T^n\hat{U}^\sigma_{z_3}\hat{U}^{\dagger\sigma}_{z_2}\}
 -N_c {\rm Tr}\{\hat{U}^\sigma_{z_1}\hat{U}^{\dagger\sigma}_{z_2}\}]
\nonumber
 \end{eqnarray}
Hereafter we use the notations  $\calz_i\equiv{(x-z_i)^2\over x_\ast}-{(y-z_i)^2\over y_\ast}$ and 
\begin{equation}
\calr~=~{(x-y)^2z_{12}^2\over x_\ast y_\ast \calz_1\calz_2}~=~{\kappa^2(\zeta_1\cdot\zeta_2)\over 2(\kappa\cdot\zeta_1)(\kappa\cdot\zeta_2)}
\end{equation}
Note that the l.h.s. of the Eq. (\ref{ope1}) is conformally invariant while the coefficient function in the r.h.s. is not.
The reason for that is the cutoff in the longitudinal direction (\ref{cutoff}). Indeed, 
we consider the light-like dipoles (in the $p_1$ direction) and impose the cutoff
 on the maximal $\alpha$ emitted by any gluon from the Wilson lines.
Formally, the light-like Wilson lines are  M\"obius invariant.
Unfortunately, the light-like Wilson lines are divergent in the longitudinal direction and moreover,  it is exactly the evolution 
equation with respect to this longitudinal cutoff which governs the high-energy behavior of amplitudes. 
At present, it is not known how to find the conformally invariant cutoff in the longitudinal direction.  When we use the non-invariant cutoff 
we expect, as usual, the invariance to hold in the leading order but to be
violated in higher orders in perturbation theory. In our calculation we restrict the longitudinal momentum of the gluons composing Wilson lines, 
and with this non-invariant cutoff the NLO evolution equation in QCD has extra non-conformal parts not related to the running of coupling constant.
Similarly, there will be non-conformal parts coming from the longitudinal cutoff of Wilson lines in the ${\cal N}=4$ SYM equation.
We will demonstrate below that it is possible to construct the
``composite conformal dipole operator'' (order by order in perturbation theory) which mimics the conformal cutoff
in the longitudinal direction so the corresponding evolution equation has no extra non-conformal parts. This is similar 
to the construction of the composite renormalized local operator in the case when the UV cutoff  does not respect the 
symmetries of the bare operator - in this case the symmetry of the UV-regularized operator is preserved 
order by order in perturbation theory by subtraction of the symmetry-restoring counterterms.
Following Ref. \cite{nlobksym} we choose the conformal composite operator in the form
\begin{eqnarray}
&&\hspace{-1mm}
[{\rm Tr}\{\hat{U}_{z_1}\hat{U}^{\dagger}_{z_2}\}\big]_{a,Y}^{\rm conf}~
\label{confodipole}\\
&&\hspace{-1mm}
=~{\rm Tr}\{\hat{U}^\sigma_{z_1}\hat{U}^{\dagger\sigma}_{z_2}\}
+{\alpha_s\over 2\pi^2}\!\int\! d^2 z_3~{z_{12}^2\over z_{13}^2z_{23}^2}
[ {\rm Tr}\{T^n\hat{U}^\sigma_{z_1}\hat{U}^{\dagger\sigma}_{z_3}T^n\hat{U}^\sigma_{z_3}\hat{U}^{\dagger\sigma}_{z_2}\}
-N_c {\rm Tr}\{\hat{U}^\sigma_{z_1}\hat{U}^{\dagger\sigma}_{z_2}\}]
\ln {4az_{12}^2\over sz_{13}^2z_{23}^2}~+~O(\alpha_s^2)
\nonumber
\end{eqnarray}
where $a$ is an arbitrary constant. It is convenient to choose the rapidity-dependent constant 
$a\rightarrow ae^{-2\eta}$ so that the 
$[{\rm Tr}\{\hat{U}^\sigma_{z_1}\hat{U}^{\dagger\sigma}_{z_2}\}\big]_a^{\rm conf}$ 
does not depend on $\eta=\ln\sigma$ and all the rapidity dependence is encoded into $a$-dependence:
\begin{eqnarray}
&&\hspace{-1mm}
[{\rm Tr}\{\hat{U}_{z_1}\hat{U}^{\dagger}_{z_2}\}\big]_a^{\rm conf}~
\label{confodipola}\\
&&\hspace{-1mm}=~{\rm Tr}\{\hat{U}^\sigma_{z_1}\hat{U}^{\dagger\sigma}_{z_2}\}
+~{\alpha_s\over 2\pi^2}\!\int\! d^2 z_3~{z_{12}^2\over z_{13}^2z_{23}^2}
[ {\rm Tr}\{T^n\hat{U}^\sigma_{z_1}\hat{U}^{\dagger\sigma}_{z_3}T^n
\hat{U}^\sigma_{z_3}\hat{U}^{\dagger\sigma}_{z_2}\}
-N_c {\rm Tr}\{\hat{U}^\sigma_{z_1}\hat{U}^{\dagger\sigma}_{z_2}\}]
\ln {4az_{12}^2\over \sigma^2 sz_{13}^2z_{23}^2}~+~O(\alpha_s^2)
\nonumber
\end{eqnarray}
Using the leading-order evolution equation \cite{npb96}
\begin{equation}
\hspace{-0mm}
{d\over d\eta}{\rm Tr}\{\hat{U}_{z_1}^\sigma\hat{U}^{\dagger\sigma}_{z_2}\}~=~\sigma{d\over d\sigma}{\rm Tr}\{\hat{U}_{z_1}^\sigma\hat{U}^{\dagger\sigma}_{z_2}\}
~=~{\alpha_s\over\pi^2}\!\int\! d^2 z_3~{z_{12}^2\over z_{13}^2z_{23}^2}[ {\rm Tr}\{T^n\hat{U}^\sigma_{z_1}\hat{U}^{\dagger\sigma}_{z_3}T^n
\hat{U}^\sigma_{z_3}\hat{U}^{\dagger\sigma}_{z_2}\}
-N_c {\rm Tr}\{\hat{U}^\sigma_{z_1}\hat{U}^{\dagger\sigma}_{z_2}\}]
\end{equation}
it is easy to see that ${d\over d\eta}[{\rm Tr}\{\hat{U}_{z_1}\hat{U}^{\dagger}_{z_2}\}\big]_a^{\rm conf}~=~0$ (with our $O(\alpha_s^2)$ accuracy). 

Rewritten in terms of conformal dipoles (\ref{confodipola}), the operator expansion (\ref{ope1})  takes the form:
\begin{eqnarray}
&&\hspace{-1mm}
(x-y)^4 T\{\halo(x)\halo^\dagger(y)\}~
=~{1\over \pi^2(N_c^2-1)}\!\int\! {d^2 z_1 d^2 z_2\over z_{12}^4}~\calr^2~\Big\{[{\rm Tr}\{\hat{U}_{z_1}\hat{U}^{\dagger}_{z_2}\}]_a^{\rm conf}
 \nonumber\\
&&\hspace{-1mm}
-~{\alpha_s\over 2\pi^2}\!\int\!  d^2z_3
{z_{12}^2\over z_{13}^2z_{23}^2}
\Big(\ln{ asz_{12}^2\over 4 z_{13}^2z_{23}^2}
\calz_3^2-i\pi+2C\Big)
[ {\rm Tr}\{T^n\hat{U}_{z_1}\hat{U}^{\dagger}_{z_3}T^n\hat{U}_{z_3}\hat{U}^{\dagger}_{z_2}\}
 -N_c {\rm Tr}\{\hat{U}_{z_1}\hat{U}^{\dagger}_{z_2}\}]_a\Big\}
 \label{opeconfa}
 \end{eqnarray}
We need to choose the new ``rapidity cutoff'' $a$ in such a way that all the energy dependence is included in the matrix element(s) of 
Wilson-line operators so the impact factor should not depend on energy ( $\equiv$ should not scale with $\rho$ as $\rho\rightarrow\infty$).  A suitable
choice of $a$ is given by $a_0=-\kappa^{-2}+i\epsilon=-{ 4x_\ast y_\ast\over s(x-y)^2}+i\epsilon$ so we obtain
\begin{eqnarray}
&&\hspace{-1mm}
(x-y)^4 T\{\halo(x)\halo^\dagger(y)\}~
=~{1\over \pi^2(N_c^2-1)}\!\int\! {d^2 z_1 d^2 z_2\over z_{12}^4}~\calr^2~\Big\{[{\rm Tr}\{\hat{U}^\sigma_{z_1}\hat{U}^{\dagger\sigma}_{z_2}\}]_{a_0}^{\rm conf}
 \nonumber\\
&&\hspace{-1mm}
-~{\alpha_s\over 2\pi^2}\!\int\!  d^2z_3
{z_{12}^2\over z_{13}^2z_{23}^2}
\Big[\ln{x_\ast y_\ast z_{12}^2\over (x-y)^2 z_{13}^2z_{23}^2}
\calz_3^2+2C\Big]
[ {\rm Tr}\{T^n\hat{U}^\sigma_{z_1}\hat{U}^{\dagger\sigma}_{z_3}
T^n\hat{U}^\sigma_{z_3}\hat{U}^{\dagger\sigma}_{z_2}\}
 -N_c {\rm Tr}\{\hat{U}^\sigma_{z_1}\hat{U}^{\dagger\sigma}_{z_2}\}]\Big\}
 \label{opeconf}
 \end{eqnarray}
where the conformal dipole $[{\rm Tr}\{\hat{U}^\sigma_{z_1}\hat{U}^{\dagger\sigma}_{z_2}\}]^{\rm conf}$ is given by Eq. (\ref{confodipola}) with
$a_0=-{4 x_\ast y_\ast\over s(x-y)^2}$.   

Now it is evident that the impact factor in the r.h.s. of this equation is M\"obius invariant and does not scale with $\rho$ so 
Eq. (\ref{confodipola}) gives conformally invariant operator up to $\alpha_s^2$ order. In higher orders, one should expect the correction 
terms with more Wilson lines. This procedure of finding the dipole with conformally regularized rapidity divergence is analogous
 to the construction of the composite renormalized local operator by adding the appropriate counterterms order by order in perturbation theory.

To find the amplitude (\ref{correl1}) in the next-to-leading order  it is sufficient to take into account only the linear evolution of Wilson-line operators
which corresponds to taking into account only two gluons in the t-channel.
The non-linear effects in the evolution (and the production) of t-channel gluons enter the four-current amplitude (\ref{correl1}) in the form of so-called
``pomeron loops'' which start from the NNLO BFKL order.
It is convenient to define the ``color dipole in the adjoint representation''
 \begin{equation}
 \hat{\cal U}^\sigma(x,y)~=~1-{1\over N_c^2-1}{\rm Tr}\{ \hat{U}^\sigma_{x} \hat{U}^{\dagger\sigma}_y\}
 \label{colodipole}
 \end{equation}
 With this two-gluon accuracy
\begin{eqnarray}
&&\hspace{-1mm}[ {1\over N_c}{\rm Tr}\{T^n\hat{U}^\sigma_{z_1}\hat{U}^{\dagger\sigma}_{z_3}T^n\hat{U}^\sigma_{z_3}\hat{U}^{\dagger\sigma}_{z_2}\}
-{\rm Tr}\{\hat{U}^\sigma_{z_1}\hat{U}^{\dagger\sigma}_{z_2}\}]~\simeq~
\nonumber\\
&&\hspace{-1mm}=~-{1\over 2}(N_c^2-1)\big[\hat{\cal U}_{\rm conf}^\sigma(z_1,z_3)+\hat{\cal U}_{\rm conf}^\sigma(z_2,z_3)-\hat{\cal U}_{\rm conf}^\sigma(z_1,z_2)\big]
\nonumber
\end{eqnarray}
The conformal dipole operator (\ref{confodipola}) in the BFKL approximation has the form:
 \begin{eqnarray}
 &&\hspace{-1mm}
 \hat{\cal U}_{\rm conf}^a(z_1,z_2)~ =~ \hat{\cal U}^\sigma(z_1,z_2)
+~{\alpha_sN_c\over 4\pi^2}\!\int\! d^2z{z_{12}^2\over z_{13}^2 z_{23}^2}\ln {4az_{12}^2\over \sigma^2 sz_{13}^2 z_{23}^2}
[\hat{\cal U}^\sigma(z_1,z_3)+\hat{\cal U}^\sigma(z_2,z_3)-\hat{\cal U}^\sigma(z_1,z_2)].
\label{caluconf}
 \end{eqnarray}
 With the two-gluon accuracy one more integration in the r.h.s. of Eq. (\ref{opeconf}) can be performed: 
\begin{eqnarray}
&&\hspace{-2mm}
{1\over \calz_2^2}\!\int\! d^2z_3{z_{23}^2\over z_{12}^2z_{13}^2\calz_3^2}\ln (a{z_{23}^2\calz_1^2\over z_{12}^2z_{13}^2})
+{1\over \calz_1^2}\!\int\! d^2z_3{z_{13}^2\over z_{12}^2z_{23}^2\calz_3^2}\ln (a{z_{13}^2\calz_2^2\over z_{12}^2z_{23}^2})
-{1\over \calz_1^2\calz_2^2}\!\int\! d^2z_3{z_{12}^2\over z_{13}^2z_{23}^2}\ln a{z_{12}^2\calz_3^2\over z_{13}^2z_{23}^2}
\nonumber\\
&&\hspace{-2mm}
=~{2\pi\over \calz_1^2\calz_2^2}\Big[
\ln {a\calz_1\calz_2\over z_{12}^2}\big(\ln {1\over  R}+{1\over R}-2\big)
+2{\rm Li}_2(1-R)-{\pi^2\over 3}-2\ln R\Big]
\end{eqnarray} 
so the resulting operator expansion takes the form
\begin{eqnarray}
&&\hspace{-1mm}      
(x-y)^4 T\{\halo(x)\halo^\dagger(y)\}~
\label{resope}\\
&&\hspace{-1mm} 
=~-{1\over\pi^2}\!\int\! {dz_1dz_2\over z_{12}^4}~
\hat{\cal U}_{\rm conf}^{a_0}(z_1,z_2)\calr^2\Big\{1-{\alpha_sN_c\over 2\pi}
\Big[\ln^2\calr-{\ln\calr\over \calr} -2C\big(\ln \calr-{1\over\calr}+2\big)+~2{\rm Li}_2(1-\calr)-{\pi^2\over 3}\Big]
\Big\}
\nonumber
 \end{eqnarray}
 We need the projection of the T-product in the l.h.s. of this equation onto the conformal eigenfunctions of the BFKL equation \cite{lip86}
\begin{equation}
\hspace{-0mm}
E_{\nu,n}(z_{10},z_{20})~
=~\Big[{\tilde{z}_{12}\over \tilde{z}_{10}\tilde{z}_{20}}\Big]^{\half+i\nu+{n\over 2}}
\Big[{\barz_{12}\over \barz_{10}\barz_{20}}\Big]^{\half+i\nu-{n\over 2}}
\label{eigenfunctions}
\end{equation}
(here $\tilde{z}=z_x+iz_y,\barz=z_x-iz_y$, $z_{10}\equiv z_1-z_0$ etc.). Since $\halo$'s are scalar operators,  the only non-vanishing contribution comes from projection on the eigenfunctions with spin $0$:
\begin{eqnarray}
&&\hspace{-2mm}
{1\over\pi^2}\!\int\! {dz_1dz_2\over z_{12}^4}~
\calr^2\Big\{1-{\alpha_sN_c\over 2\pi}
\Big[\ln^2\calr-{\ln\calr\over \calr} -2C\big(\ln \calr-{1\over\calr}+2\big)+~2{\rm Li}_2(1-\calr)-{\pi^2\over 3}\Big]
\Big\}\Big[{z_{12}^2\over z_{10}^2z_{20}^2}\Big]^{\half+i\nu}
\nonumber\\
&&\hspace{-2mm}
=~\Big[{\kappa^2\over (2\kappa\cdot\zeta_0)^2}\Big]^{\half+i\nu} {\Gamma^2\big(\half-i\nu\big)\over \Gamma (1-2i\nu)}
{\big({1\over 4}+\nu^2\big)\pi\over \cosh\pi\nu}
\Big\{1+{\alpha_sN_c\over 2\pi}\Phi_1(\nu)\Big\}
\label{fla26}
\end{eqnarray} 
where  
\begin{equation}
\hspace{-0mm}
\Phi_1(\nu)~=~-2\psi'\big(\half+i\nu\big)-2\psi'\big(\half-i\nu\big)+{2\pi^2\over 3}
+{\chi(\nu)-2\over \nu^2+{1\over 4}}+2C\chi(\nu)
\label{Fi}
\end{equation}
and
$\zeta_0\equiv p_1+{z_{0\perp}^2\over s}p_2+z_{0\perp}$.

Now, using the decomposition of the product of the transverse $\delta$-functions in conformal 3-point functions $E_{\nu,n}(z_{10},z_{20})$
\begin{equation}
\hspace{-1mm}
\delta^{(2)}(z_1-w_1)\delta^{(2)}(z_2-w_2)~
=~\sum_{n=-\infty}^\infty\!\int\! {d\nu\over \pi^4}~{\nu^2+{n^2\over 4}\over z_{12}^2w_{12}^2}
\int\! d^2\rho~ E^\ast_{\nu,n}(w_1-\rho,w_2-\rho)E_{\nu,n}(z_1-\rho,z_2-\rho)
\label{lobzor120}
\end{equation}
we obtain 
\begin{eqnarray}
&&\hspace{-1mm}      
(x-y)^4 T\{\halo(x)\halo^\dagger(y)\}~
\nonumber\\
&&\hspace{-1mm}=~-\int\! d\nu\!\int\! d^2z_0~{\nu^2(1+4\nu^2)\over 4\pi\cosh\pi\nu}
{\Gamma^2\big(\half-i\nu\big)\over\Gamma(1-2i\nu)}
~\Big({\kappa^2\over 4(\kappa\cdot\zeta_0)^2}\Big)^{\half +i\nu}\big[1+{\alpha_sN_c\over 2\pi}\Phi_1(\nu)\big]
 \hat{\cal U}_{\rm conf}^a(\nu,z_0)
 \label{opeu}
 \end{eqnarray}
where 
\begin{eqnarray}
&&\hspace{-1mm}
 \hat{\cal U}_{\rm conf}^a(\nu,z_0)~\equiv~{1\over \pi^2}\!\int\! {d^2z_1d^2z_2\over z_{12}^4}~
\Big({z_{12}^2\over z_{10}^2z_{20}^2}\Big)^{\half -i\nu}~ \hat{\cal U}_{\rm conf}^a(z_1,z_2)
\label{opconfspin}
\end{eqnarray}
is a conformal dipole in the $z_0,\nu$ representation.

Similarly, one can write down the expansion of the bottom part of the diagram in color dipoles:
\begin{eqnarray}
&&\hspace{-1mm}      
(x'-y')^4 T\{\halo(x')\halo^\dagger(y')\}~
\nonumber\\
&&\hspace{-1mm}
=~-\!\int\! d\nu'\!\int\! d^2z'_0~{{\nu'}^2(1+4{\nu'}^2)\over 4\pi\cosh\pi\nu'}
{\Gamma^2\big(\half-i\nu'\big)\over\Gamma(1-2i\nu')}
~\Big({{\kappa'}^2\over 4(\kappa'\cdot\zeta'_0)^2}\Big)^{\half +i\nu'}\big[1+{\alpha_sN_c\over 2\pi}\Phi_1(\nu')\big]
 \hat{\cal V}_{\rm conf}^{b_0}(\nu',z'_0).
 \label{opev}
 \end{eqnarray}
Here $\zeta'_0\equiv p_1+{{z'}_{0\perp}^2\over s}p_2+z'_{0\perp}$,  
$b_0=-{\kappa'}^{-2}=-{4 x'_\bu y'_\bu\over s(x'-y')^2}+i\epsilon$, and
\begin{equation}
 \hat{\cal V}_{\rm conf}^{b}(\nu',z'_0)~=~{1\over\pi^2}\!\int\! {d^2z_1d^2z_2\over z_{12}^4}~
\Big({z_{12}^2\over z_{10}^2z_{20}^2}\Big)^{\half -i\nu'}~ \hat{\cal V}_{\rm conf}^b(z_1,z_2),
\label{vopconfspin}
\end{equation}
where the conformal operator
\begin{eqnarray}
&&\hspace{-2mm} 
 \hat{\cal V}_{\rm conf}^b(z_1,z_2)~
 =~ \hat{\cal V}^\sigma(z_1,z_2)+{\alpha_sN_c\over 4\pi^2}\!\int\! d^2z{z_{12}^2\over z_{13}^2 z_{23}^2}
\ln {4b\sigma^2z_{12}^2\over sz_{13}^2 z_{23}^2}
[\hat{\cal V}^\sigma(z_1,z_3)+\hat{\cal V}^\sigma(z_2,z_3)-\hat{\cal V}^\sigma(z_1,z_2)]
\label{calvconf}
\end{eqnarray}
 is made from the dipoles  
$
 \hat{\cal V}^\sigma(x,y)~=~1-{1\over N_c^2-1}{\rm Tr}\{ \hat{V}^\sigma_{x} \hat{V}^{\dagger\sigma}_y\}
 $
(cf. Eq. (\ref{colodipole}))  ordered along the straight line $\parallel~p_2$ with the rapidity restriction
\begin{eqnarray}
&&\hspace{-2mm} 
 V^\sigma_x~=~{\rm Pexp}\Big[ig\!\int_{-\infty}^\infty\!\! du~ p_1^\mu A^\sigma_\mu(up_2+x_\perp)\Big]
 \nonumber\\ 
&&\hspace{-2mm}
A^\sigma_\mu(x)~=~\int\! d^4 k ~\theta(\sigma-|\beta_k|)e^{-ik\cdot x} A_\mu(k)
\label{cutoffv}
\end{eqnarray}

If we substitute both operator expansions (\ref{opeu}) and (\ref{opev}) into the correlation function (\ref{correl1}), it takes the form
\begin{eqnarray}
&&\hspace{-1mm}      
(x-y)^4 (x'-y')^4\langle T\{\halo(x)\halo^\dagger(y)\halo(x')\halo^\dagger(y')\}\rangle~
\label{fla35}\\
&&\hspace{-1mm}
=~\!\int\! d\nu d\nu'\!\int\! d^2z_0d^2z'_0~{\nu^2(1+4\nu^2)\over 4\pi\cosh\pi\nu}
{\Gamma^2\big(\half-i\nu\big)\over\Gamma(1-2i\nu)}\Big({\kappa^2\over 4(\kappa\cdot\zeta_0)^2}\Big)^{\half +i\nu}\big[1+{\alpha_sN_c\over 2\pi}\Phi_1(\nu)\big]
\nonumber\\
&&\hspace{-1mm}
\times~
{{\nu'}^2(1+4{\nu'}^2)\over 4\pi\cosh\pi\nu'}
{\Gamma^2\big(\half-i\nu'\big)\over\Gamma(1-2i\nu')}
\Big({{\kappa'}^2\over 4(\kappa'\cdot\zeta'_0)^2}\Big)^{\half +i\nu'}
\big[1+{\alpha_sN_c\over 2\pi}\Phi_1(\nu')\big]
\langle \hat{\cal U}_{\rm conf}^{a_0}(\nu,z_0) \hat{\cal V}_{\rm conf}^{b_0}(\nu',z'_0)\rangle
\nonumber
  \end{eqnarray}
%

\section{NLO scattering of conformal dipoles and the NLO amplitude}
The last step is to find the NLO amplitude of the  scattering of conformal dipoles $\hat{\cal U}^a_{\rm conf}(z_0,\nu)$ and $\hat{\cal V}^b_{\rm conf}(z'_0,\nu')$. 
First we need to write down the 
 NLO BFKL evolution (as we discussed above the rapidity dependence is now encoded in the $a$-evolution):
\begin{eqnarray}
&&\hspace{-2mm}
2a{d\over da}\hat{\cal U}^a_{\rm conf}(z_1,z_2)~
~=~
\!\int\!d^2z_3d^2z_4 K(z_1,z_2;z_3,z_4)~
\hat{\cal U}^a_{\rm conf}(z_3,z_4)
\label{nlobfklconf}
\end{eqnarray}
where the kernel $K(z_1,z_2;z_3,z_4)$ in the first two orders has the form\cite{nlobk, nlobfklconf}
\begin{eqnarray}
&&\hspace{-4mm}
K_{\rm LO}(z_1,z_2;z_3,z_4)~=~{\alpha_sN_c\over 2\pi^2}\Big[{z_{12}^2\delta^{2}(z_{13})\over z_{14}^2z_{24}^2}
+{z_{12}^2\delta^{2}(z_{24})\over z_{13}^2z_{23}^2}
-~\delta^{2}(z_{13})\delta^{2}(z_{24})\!\int\! d^2z~{z_{12}^2\over (z_1-z)^2(z_2-z)^2}\Big]
\label{klo}
\end{eqnarray}
\begin{eqnarray}
&&\hspace{-3mm}
K_{\rm NLO}(z_1,z_2;z_3,z_4)~=~-{\alpha_sN_c\over 4\pi}{\pi^2\over 3}K_{\rm LO}(z_1,z_2;z_3,z_4)
\nonumber\\
&&\hspace{-3mm}
+~{\alpha_s^2N_c^2\over 8\pi^4 z_{34}^4}~\Bigg[{z_{12}^2z_{34}^2\over z_{13}^2z_{24}^2}
\Big\{\Big(
1+{z_{12}^2z_{34}^2\over z_{13}^2z_{24}^2- z_{14}^2z_{23}^2}\Big)
\ln{z_{13}^2z_{24}^2\over z_{14}^2z_{23}^2}
+2\ln{z_{12}^2z_{34}^2\over z_{14}^2z_{23}^2}\Big\}
+12\pi^2\zeta(3)z_{34}^4\delta(z_{13})\delta(z_{24})
\Bigg]
\label{nlokonf}
\end{eqnarray}
The eigenfunctions of the kernel $K$ are given by Eq. (\ref{eigenfunctions})
and the eigenvalues by the pomeron intercept (\ref{eigen1}).
\begin{equation}
\int\!d^2z_3d^2z_4 ~K(z_1,z_2;z_3,z_4) E_{\nu,n}(z_{30},z_{40})~=~\omega(n,\nu)E_{\nu,n}(z_{10},z_{20})
\end{equation}
For the composite operators with definite conformal spin (\ref{opconfspin}) the 
evolution equation (\ref{nlobfklconf})  takes the simple form
\begin{equation}
\hspace{-0mm}
2a{d\over da} \hat{\cal U}_{\rm conf}^a(\nu,z_0)~=~\omega(\nu) \hat{\cal U}_{\rm conf}^a(\nu,z_0)
\label{evolv}
\end{equation}
(Since Eq. (\ref{correl1}) is a correlation functions of scalar currents,  we need only the $n=0$ projection of this evolution).

The result of the evolution is 
\begin{equation}
\hspace{-0mm}
\hat{\cal U}_{\rm conf}^a(\nu,z_0)~=~(a/\tilde{a})^{\half\omega(\nu)} \hat{\cal U}_{\rm conf}^{a_0}(\nu,z_0)
\label{evolresult}
\end{equation}
where the endpoint of the evolution $\tilde{a}$ should be taken from the requirement that the amplitude of scattering of conformal dipoles
with ``normalization points'' $\tilde{a}$ and $b$ should not contain large logarithms of energy so it will serve as the initial point
of the evolution. (This is similar to taking $\mu^2$ around 1 GeV for the initial point of the DGLAP evolution).  The amplitude of scattering of two conformal dipoles
is calculated in the Appendix and the result has the form (see Eq. (\ref{fla57})):
\begin{eqnarray}
&&\hspace{-5mm}
\langle\hat{\cal U}_{\rm conf}^{\tilde a}(\nu,z_0)\hat{\cal V}_{\rm conf}^b(\nu',z'_0)\rangle~
=~-{\alpha_s^2N_c^2\over N_c^2-1}
{16\pi^2\over \nu^2(1+4\nu^2)^2}
\Big[\delta(z_0-z'_0)\delta(\nu+\nu')
\label{confdipscatlow}\\
&&\hspace{-5mm}
+{2^{1-4i\nu}\delta(\nu-\nu')\over \pi|z_0-z'_0|^{2-4i\nu}}
{\Gamma\big({1\over 2}+i\nu\big)\Gamma(1-i\nu)\over\Gamma(i\nu)\Gamma\big(\half-i\nu\big)}\Big]\Big[1+{\alpha_sN_c\over 2\pi}\Big(\chi(\nu)\big[\ln \tilde{a}b -i\pi-4C
-{2\over\nu^2+{1\over 4}}\big]-{\pi^2\over 3}\Big)\Big].
\nonumber
 \end{eqnarray}
Using Eq. (\ref{confdipscatlow})  as an initial condition for the evolution (\ref{evolresult}) we get the following amplitude of scattering of two conformal dipoles:
\begin{eqnarray}
&&\hspace{-1mm}
\langle\hat{\cal U}_{\rm conf}^a(\nu,z_0)\hat{\cal V}_{\rm conf}^b(\nu',z'_0)\rangle~
\label{confdipscat}\\
&&\hspace{-1mm}
=~i{\alpha_s^2N_c^2\over N_c^2-1}
{(a+i\epsilon)^{\omega(\nu)\over 2}(b+i\epsilon)^{\omega(\nu)\over 2}-(-a-i\epsilon)^{\omega(\nu)\over 2}(-b-i\epsilon)^{\omega(\nu)\over 2}\over\sin\pi\omega(\nu)}
\Big[1-{\alpha_sN_c\over 2\pi}\Big(\chi(\gamma)\big[4C+{8\over 1+4\nu^2)}\big]+{\pi^2\over 3}\Big)\Big]
\nonumber\\
&&\hspace{-3mm}
\times~{16\pi^2\over \nu^2(1+4\nu^2)^2}
\Big[\delta(z_0-z'_0)\delta(\nu+\nu')+{2^{1-4i\nu}\delta(\nu-\nu')\over \pi|z_0-z'_0|^{2-4i\nu}}
{\Gamma\big({1\over 2}+i\nu\big)\Gamma(1-i\nu)\over\Gamma(i\nu)\Gamma\big(\half-i\nu\big)}\Big].
\nonumber
 \end{eqnarray}
Finally, substituting this amplitude in Eq. (\ref{fla35}) we obtain
\begin{eqnarray}
&&\hspace{-1mm}      
(x-y)^4 (x'-y')^4\langle T\{\halo(x)\halo^\dagger(y)\halo(x')\halo^\dagger(y')\}\rangle~
\nonumber\\
&&\hspace{-1mm}
=~i{\alpha_s^2N_c^2\over N_c^2-1}\!\int\! d\nu 
{(a_0+i\epsilon)^{\omega(\nu)\over 2}(b_0+i\epsilon)^{\omega(\nu)\over 2}-(-a_0-i\epsilon)^{\omega(\nu)\over 2}(-b_0-i\epsilon)^{\omega(\nu)\over 2}\over\sin\pi\omega(\nu)}
~\Big\{1-{\alpha_sN_c\over 2\pi}\Big(\chi(\nu)\Big[4C+{8\over 1+4\nu^2}\Big]+{\pi^2\over 3}\Big)\Big\}
\nonumber\\
&&\hspace{-1mm}
\times~2\pi\nu{\tanh\pi\nu\over \cosh^2\pi\nu}\!\int\! d^2z_0~\Big({\kappa^2\over 4(\kappa\cdot\zeta_0)^2}\Big)^{\half +i\nu}
\Big({{\kappa'}^2\over 4(\kappa'\cdot\zeta'_0)^2}\Big)^{\half -i\nu}
\big[1+{\alpha_sN_c\over 2\pi}\Phi_1(\nu)\big]\big[1+{\alpha_sN_c\over 2\pi}\Phi_1(\nu)\big]
\label{result}
  \end{eqnarray}
where we used the integral
\begin{eqnarray}
&&\hspace{-2mm}
\int\! {d^2z'_0\over [(z_0-z'_0)^2]^{1-2i\nu}}
\Big[{{\kappa'}^2\over 4(\kappa'\cdot\zeta'_0)^2}\Big]^{\half+i\nu}
~=~{\pi\over 2i\nu}\Big[{{\kappa'}^2\over 4(\kappa'\cdot\zeta_0)^2}\Big]^{\half-i\nu}
\end{eqnarray}
Now it is easy to see that Eq. (\ref{result})  coincides with Eq. (\ref{koppinkoop}) with 
\begin{eqnarray}
&&\hspace{-0mm}
F(\nu)~=~{N_c^2\over N_c^2-1}{4\pi^4\alpha_s^2\over\cosh^2\pi\nu}~
\big\{1+{\alpha_sN_c\over 2\pi}\Phi_1(\nu)\big]^2\big\}~
\Big\{1-{\alpha_sN_c\over 2\pi}\Big[\chi(\nu)\Big(4C+{8\over 1+4\nu^2}\Big)
+{\pi^2\over 3}\Big]+O(\alpha_s^2)\Big\}
\nonumber\\
&&\hspace{-0mm}
=~{N_c^2\over N_c^2-1}{4\pi^4\alpha_s^2\over\cosh^2\pi\nu}~
\Big\{1+{\alpha_sN_c\over \pi}\Big[-2\psi'\big(\half+i\nu\big)-2\psi'\big(\half-i\nu\big)
+{\pi^2\over 2}-{8\over 1+4\nu^2}\Big]+O(\alpha_s^2)\Big\}
\end{eqnarray}
which gives the pomeron residue in the next-to-leading order(recall that $a_0=-\kappa^{-2}+i\epsilon$ and $b_0=-{\kappa'}^2+i\epsilon$). 
The lowest-order term in this formula agrees with the leading-order impact factor (\ref{loif}) 
calculated in Refs. \cite{penecostalba,penecostalba2}.

\section{Conclusions}
The main result of the paper is that the rapidity  factorization and high-energy operator expansion in color dipoles works at the NLO level. 
There are many examples of the factorization which are fine at the leading order but fail at the NLO level. We believe that the high-energy
OPE has the same status as usual light-cone expansion in light-ray operators so one can calculate the high-energy amplitudes level by level in 
perturbation theory. 
As an outlook we intend to apply the NLO high-energy operator expansion for the description of QCD amplitudes.  Although our composite dipole 
(\ref{confodipola}) is no longer conformal in QCD, we believe that the effects due to the running coupling 
calculated in Refs. \cite{prd75,kw1} can be incorporated in some sort of structure resembling the formula (\ref{koppinkoop}) for ${\cal N}=4$
SYM.
As an application of the machinery developed here we intend to calculate the photon impact factor for the 
structure function $F_2(x)$ of deep inelastic scattering which will compete the calculation of  small-$x$ structure functions at the NLO level.
The study is in progress.

The authors are grateful to  L.N. Lipatov and J. Penedones for valuable discussions. 
This work was supported by contract
 DE-AC05-06OR23177 under which the Jefferson Science Associates, LLC operate the Thomas Jefferson National Accelerator Facility.
The work of G.A.C is supported by the grant ANR-06-JCJC-0084
\section{Appendix. Dipole-dipole scattering in the NLO}
The amplitude of scattering of two conformal dipoles in the first two orders of perturbation theory can be easily calculated in the momentum representation.  
In the leading order it has the form
\begin{eqnarray}
&&\hspace{-1mm}
\langle\hat{\cal U}(z_1,z_2)\hat{\cal V}(z'_1,z'_2)\rangle~=~-{\alpha_s^2N_c^2\over 2\pi^2(N_c^2-1)}\!\int\! {d^2k_\perp d^2q_\perp\over k_\perp^2(q-k)_\perp^2}
\big[e^{i(k,z_1)_\perp}-e^{i(k,z_2)_\perp}\big]\big[e^{i(q-k,z_1)_\perp}-e^{i(q-k,z_2)_\perp}\big]
\nonumber\\
&&\hspace{-1mm}
\times~
\big[e^{-i(k,z'_1)_\perp}-e^{-i(k,z'_2)_\perp}\big]\big[e^{-i(q-k,z'_1)_\perp}-e^{-i(q-k,z'_2)_\perp}\big]
~=~-{\alpha_s^2N_c^2\over 2(N_c^2-1)}\ln^2{(z_1-z'_1)_\perp^2(z_2-z'_2)_\perp^2\over (z_1-z'_2)_\perp^2(z_2-z'_1)_\perp^2}
\label{ddlorder}
\end{eqnarray}
The dipole-dipole amplitude in the next-to-leading order can be taken from Ref.  \cite{babbal}
\begin{eqnarray}
&&\hspace{-1mm}           
\langle\hat{\cal U}(z_1,z_2)\hat{\cal V}(z'_1,z'_2)\rangle~
\nonumber\\
&&\hspace{-1mm}
=~i{\alpha_s^3N_c^3s\over 4\pi^5(N_c^2-1)}
\!\int\! {d^2k_\perp d^2k'_\perp d^2q_\perp \over k_\perp^2(q-k)_\perp^2}\big[e^{i(k,z_1)_\perp}-e^{i(k,z_2)_\perp}\big]\big[e^{i(q-k,z_1)_\perp}-e^{i(q-k,z_2)_\perp}\big]
\!\int\! {d\alpha d\beta\over \alpha\beta s-(k-k')_\perp^2+i\epsilon}
\nonumber\\
&&\hspace{-1mm}
\times~
~\Bigg(
{q_\perp^2-{1\over\alpha\beta s}[k_\perp^2(q-k')_\perp^2+{k'}_\perp^2(q-k)_\perp^2]\over {k'}_\perp^2(q-k')_\perp^2}
\big[e^{-i(k',z'_1)_\perp}-e^{-i(k',z'_2)_\perp}\big]\big[e^{-i(q-k',z'_1)_\perp}-e^{-i(q-k',z'_2)_\perp}\big]           
\nonumber\\
&&\hspace{-1mm}
+~{1\over 2\alpha\beta s}\Big[{k_\perp^{2}\over \alpha\beta s-{k'}_\perp^2+i\epsilon}
+{(q-k)_\perp^{2}\over \alpha\beta s-(q-k')_\perp^2+i\epsilon}\Big]
\big[e^{-i(k,z'_1)_\perp}-e^{-i(k,z'_2)_\perp}\big]\big[e^{-i(q-k,z'_1)_\perp}-e^{-i(q-k,z'_2)_\perp}\big]
\Bigg)
\label{ddnlorder}
\end{eqnarray}
where the singularities $1/\alpha$ and $1/\beta$ are understood in a principal value sense. Imposing the  ``rigid cutoffs''  $|\alpha|<\sigma$,  $|\beta|<\sigma'$ 
(in contrast to ``slope'' cutoffs used in Ref. \cite{babbal}) and using the integrals
\begin{equation}
\hspace{-0mm}
s\!\int_{-\sigma}^{\sigma}\! d\alpha\! \int_{-\sigma'}^{\sigma'}\! d\beta~{1\over\alpha\beta s-k_\perp^2+i\epsilon}~=~
\!\int_{-\sigma}^{\sigma}d\alpha\! \int_{-\sigma'}^{\sigma'} \! d\beta~{k_\perp^2\over\alpha\beta s-k_\perp^2+i\epsilon}\big(P{1\over\alpha}\big)\big(P{1\over\beta}\big)
~=~-2i\pi\big(\ln{\sigma\sigma' s\over k_\perp^2}-{i\pi\over 2}\big)+O\big({k_\perp^2\over s}\big)
\end{equation}
we obtain
\begin{eqnarray}
&&\hspace{-2mm}
\langle\hat{\cal U}^{\sigma}(z_1,z_2)\hat{\cal V}^{\sigma'}(z'_1,z'_2)\rangle~=~-{\alpha_s^3N_c^3\over 2\pi^4(N_c^2-1)}F^{\sigma,\sigma'}(z_1,z_2;z'_1,z'_2),
\label{fla45}
\end{eqnarray}
where
\begin{eqnarray}
&&\hspace{-2mm}
F^{\sigma,\sigma'}(z_1,z_2;z'_1,z'_2)~=~\!\int\!d^2k d^2k' d^2q (e^{i(k,z_1)}-e^{i(k,z_2)})(e^{i(q-k,z_1)}-e^{i(q-k,z_2)})
{1\over k^2(q-k)^2} 
 \label{fla46}\\
&&\hspace{-2mm}
\Bigg(\Big[{k^2\over  {k'}^2(k-k')^2}
+{(q-k)^2\over (q-k')^2(k-k')^2}-{q^2\over {k'}^2(q-k')^2}\Big]{\ln {\sigma_1\sigma_2s\over(k-k')^2}-{i\pi\over 2}\over (k-k')^2}
 (e^{-i(k',z'_1)}-e^{-i(k',z'_2)})
 \nonumber\\
&&\hspace{-2mm}
\times~(e^{-i(q-k',z'_1)}-e^{-i(q-k',z'_2)})-~\Big({k^2/2\over {k'}^2-(k-k')^2}\Big[
{\ln {\sigma_1\sigma_2s\over(k-k')^2}-{i\pi\over 2}\over (k-k')^2}-{\ln {\sigma_1\sigma_2s\over {k'}^2}-{i\pi\over 2}\over {k'}^2}\Big]
 \nonumber\\
&&\hspace{-2mm}
-~{(q-k)^2/2\over (q-k')^2-(k-k')^2}\Big[
{\ln {\sigma_1\sigma_2s\over(k-k')^2}-{i\pi\over 2}\over (k-k')^2}-{\ln {\sigma_1\sigma_2s\over(q-k')^2}-{i\pi\over 2}\over (q-k')^2}\Big]\Big) (e^{-i(k,z'_1)}-e^{-i(k,z'_2)})(e^{-i(q-k,z'_1)}-e^{-i(q-k,z'_2)})\Bigg)
\nonumber 
\end{eqnarray}
Adding the ``correction terms''   (\ref{caluconf}) and (\ref{calvconf}) which make the dipoles  conformal, we get
\begin{eqnarray}
&&\hspace{-1mm}
-{2\over \alpha_s^2}{N_c^2-1\over N_c^2}\langle [\calu(z_1,z_2)]_{\tilde a}^{\rm conf}[\calv(z'_1,z'_2)]_b^{\rm conf}\rangle
~=~\ln^2{z_{1'}^2z_{22'}^2\over z_{12'}^2 z_{21'}^2}+{\alpha_sN_c\over\pi^4}F^{\sigma,\sigma'}(z_1,z_2;z'_1,z'_2)
\nonumber\\
&&\hspace{-1mm}
+~
{\alpha_sN_c\over 4\pi^2}\!\int\! d^2z_3{z_{12}^2\over z_{13}^2 z_{23}^2}\ln {4\tilde{a}z_{12}^2\over \sigma^2sz_{13}^2 z_{23}^2}
[\ln^2\{z_1,z_3;z'_1,z'_2\}+\ln^2\{z_2,z_3;z'_1,z'_2\}-\ln^2\{z_1,z_2;z'_1,z'_2\}]
\nonumber\\
&&\hspace{-1mm}
+~{\alpha_sN_c\over 4\pi^2}\!\int\! d^2z'_3{ z_{1'2'}^2\over z_{1'3'}^2 z_{2'3'}^2}\ln {4bz_{1'2'}^2\over {\sigma'}^2sz_{1'3'}^2 z_{2'3'}^2}
[\ln^2\{z'_1,z'_3;z_1,z_2\}+\ln^2\{z'_2,z'_3;z_1,z_2\}-\ln^2\{z'_1,z'_2;z_1,z_2\}]
 \label{fla47}
 \end{eqnarray}
where $\ln\{z_1,z_2;z'_1,z'_2\}\equiv\ln^2{(z_1-z'_1)_\perp^2(z_2-z'_2)_\perp^2\over (z_1-z'_2)_\perp^2(z_2-z'_1)_\perp^2}$.
Unfortunately, we were not able to perform the Fourier transformation in Eq. (\ref{fla46}) explicitly. However,
as discussed above, we need only the projection of this amplitude on the 
eigenfunctions with conformal spin 0 which can be easily calculated. Taking $z_0=0$ and performing inversion we get
\begin{equation}
\hspace{1mm}
\int\! {d^2z'_1d^2z'_2\over {z'}_{12}^4}\langle \calu^{\tilde{a}}_{\rm conf}(z_1,z_2)
\calv^b_{\rm conf}(z'_1,z'_2)\rangle
\Big({z_{1'2'}^2\over {z'}_{10}^2{z'}_{20}^2}\Big)^{\half+i\nu}~\rightarrow~\int\! d^2z'_1d^2z'_2\langle \calu^{\tilde{a}}_{\rm conf}(z_1,z_2)
\calv^b_{\rm conf}(z'_1,z'_2)\rangle
\big(z_{1'2'}^2\big)^{-{3\over 2}+i\nu}
 \end{equation}
The r.h.s. of this equation corresponds to forward scattering and can be easily calculated. Using 
$(z^2)^{\gamma-2}={4^{\gamma-2}\over\pi \Gamma(2-\gamma)}\!\int\! d^2p~(p^2)^{1-\gamma}\Gamma(\gamma-1)e^{i(p,x)_\perp}$
one obtains
\begin{equation}
\hspace{-0mm}
\int\! d^2z'_1d^2z'_2~F^{\sigma,\sigma'}(z_1,z_2;z'_1,z'_2)(z_{1'2'}^2)^{\gamma-2}
~=~{2\pi^5(z_{12}^2)^\gamma\over \gamma^2(1-\gamma)^2}
\Big\{\bar\chi(\gamma)\big[\ln {\sigma^2{\sigma'}^2s^2\over 16}z_{12}^4-i\pi+2\bar{\chi}(-\gamma)-4C\big]+\bar{\chi}'(\gamma)-\bar\chi^2(\gamma)-{\pi^2\over 3}\Big\}
\label{fla49}
 \end{equation}
where $\bar\chi(\gamma)=\chi\big(i(\half-\gamma)\big)=2C-\psi(\gamma)-\psi(1-\gamma)$. The integration of the ``correction terms'' can be performed in the coordinate space and the result is
\begin{eqnarray}
&&\hspace{-1mm}
{\alpha_sN_c\over 4\pi^2}\int\! d^2z'_1d^2z'_2~
(z_{1'2'}^2)^{\gamma-2}
\Big\{\!\int\! d^2z_3{z_{12}^2\over z_{13}^2 z_{23}^2}\ln {4\tilde{a}z_{12}^2\over \sigma^2sz_{13}^2 z_{23}^2}
[\ln^2\{z_1,z_3;z'_1,z'_2\}+\ln^2\{z_2,z_3;z'_1,z'_2\}-\ln^2\{z_1,z_2;z'_1,z'_2\}]\nonumber\\
&&\hspace{-1mm}
+~\!\int\! d^2z'_3{ z_{1'2'}^2\over z_{1'3'}^2 z_{2'3'}^2}\ln {4bz_{1'2'}^2\over {\sigma'}^2sz_{1'3'}^2 z_{2'3'}^2}
[\ln^2\{z_1,z_3;z'_1,z'_2\}+\ln^2\{z_2,z_3;z'_1,z'_2\}-\ln^2\{z_1,z_2;z'_1,z'_2\}]\Big\}
\nonumber\\
&&\hspace{-1mm}
=~{\alpha_sN_c\pi\over\gamma^2(1-\gamma)^2}(z_{12}^2)^{\gamma}
\Big[2\bar\chi(\gamma)[-\ln{\sigma^2{\sigma'}^2s^2z_{12}^4\over 16\tilde{a}b}+\bar\chi(\gamma-1)-\bar\chi(-\gamma)\big]-2\bar\chi^2(\gamma)-2\chi'(\gamma)\Big]
 \label{corresult}
 \end{eqnarray}
Adding Eqs. (\ref{fla49}) and (\ref{corresult}), making another inversion and restoring $z_0$ we get
\begin{eqnarray}
&&\hspace{-1mm}
\int\! {d^2z'_1d^2z'_2\over {z'}_{12}^4}\langle \calu^{\tilde{a}}_{\rm conf}(z_1,z_2)
\calv^b_{\rm conf}(z'_1,z'_2)\rangle
\Big[{z_{1'2'}^2\over {z'}_{10}^2{z'}_{20}^2}\Big]^{\half+i\nu}
\nonumber\\
&&\hspace{-1mm}
=~-2\pi^2\alpha_s^2{N_c^2\over N_c^2-1}{16\over (1+4\nu^2)^2}\Big[{z_{12}^2\over z_{10}^2z_{20}^2}\Big]^{\half+i\nu}
~\Big\{1+{\alpha_sN_c\over 2\pi}\Big(\chi(\nu)\big[\ln \tilde{a}b-i\pi-4C-{2\over\nu^2+{1\over 4}}\big]-{\pi^2\over 3}\Big)\Big\}
 \end{eqnarray}
so the lowest-order amplitude of scattering of two conformal dipoles   is
\begin{eqnarray}
&&\hspace{-5mm}
\langle\hat{\cal U}_{\rm conf}^{\tilde a}(\nu,z_0)\hat{\cal V}_{\rm conf}^b(\nu',z'_0)\rangle~
=~-{\alpha_s^2N_c^2\over N_c^2-1}
{16\pi^2\over \nu^2(1+4\nu^2)^2}
\Big[\delta(z_0-z'_0)\delta(\nu+\nu')
\label{fla57}\\
&&\hspace{-5mm}
+~{2^{1-4i\nu}\delta(\nu-\nu')\over \pi|z_0-z'_0|^{2-4i\nu}}
{\Gamma\big({1\over 2}+i\nu\big)\Gamma(1-i\nu)\over\Gamma(i\nu)\Gamma\big(\half-i\nu\big)}\Big]\Big[1+{\alpha_sN_c\over 2\pi}\Big(\chi(\nu)\big[\ln \tilde{a}b -i\pi-4C
-{2\over\nu^2+{1\over 4}}\big]-{\pi^2\over 3}\Big)\Big]
\nonumber
 \end{eqnarray}
where we used the orthogonality condition \cite{lip86} for the eigenfunctions (\ref{eigenfunctions})
\begin{eqnarray}
&&\hspace{-1mm}
\int\! {d^2z_1 d^2z_2\over z_{12}^4}~ E^\ast_{\nu',m}(z_1-z'_0,z_2-z'_0)E_{\nu,n}(z_1-z_0,z_2-z_0)~=~{\pi^4\over 2\big(\nu^2+{n^2\over 4}\big)}\Bigg[\delta(\nu-\nu')\delta_{m,n}\delta^{(2)}(z_0-z'_0)
\label{ortho}\\
&&\hspace{-3mm}
+~\delta(\nu+\nu')\delta_{m,-n}(\tilde{z}_0-\tilde{z}'_0)^{-1+n-2i\nu}(\bar{z}_0-\bar{z}'_0)^{-1-n-2i\nu}{2^{4i\nu+1}\over \pi}
\Big({|n|\over 2}+i\nu\Big){\Gamma\big({1+|n|\over 2}-i\nu\big)\Gamma\big({|n|\over 2}+i\nu\big)\over \Gamma\big({1+|n|\over 2}+i\nu\big)\Gamma\big({|n|\over 2}-i\nu\big)}\Bigg]
\nonumber
 \end{eqnarray}

\end{document}